\documentclass{article}
\usepackage{arxiv}
\usepackage{amsmath}

\usepackage[utf8]{inputenc} 
\usepackage[T1]{fontenc}    
\usepackage{hyperref}       
\usepackage{url}            
\usepackage{booktabs}       
\usepackage{amsfonts}       
\usepackage{nicefrac}       
\usepackage{microtype}      
\usepackage{lipsum}		
\usepackage{graphicx}
\usepackage{doi}
\usepackage{xcolor}

\usepackage{csquotes}
\usepackage[backend=biber]{biblatex}

\title{Stochastic Distribution Network Reconfiguration under Load Uncertainty}

\date{October 7, 2026}	

\author{
Celso Cavellucci \qquad Fábio Luiz Usberti \\
Institute of Computing, University of Campinas \\
Av. Albert Einstein, 1251, 13083-852 Campinas, SP, Brazil
}

\renewcommand{\shorttitle}{Stochastic Reconfiguration under Load Uncertainty}

\renewenvironment{abstract}
  {\centerline{\large\bfseries Abstract}
   \begin{quote}}
  {\par
   \end{quote}}

\renewcommand{\keywords}[1]{%
  \par\noindent\textbf{Keywords:} #1%
}

\renewcommand{\headeright}{Technical Report}
\renewcommand{\undertitle}{Technical Report}

\begin{document}

\maketitle

\begin{abstract}
This paper investigates distribution network reconfiguration under demand uncertainty using a two-stage stochastic formulation. The network topology is selected in the first stage, whereas the electrical variables are determined separately for each scenario. The deterministic equivalent is formulated as a MISOCP problem that accounts for power balances, losses, voltage constraints, and radiality. Experiments on 12 networks, ranging from 17 to 10\,561 nodes, consider three demand scenarios. Reconfiguration reduced expected losses in all instances, with reductions ranging from $3.23\%$ to $65.23\%$ and averaging $31.76\%$, while also reducing the number of voltage violations in the \textit{ex post} evaluation by approximately $55\%$. Eleven of the 12 problems were solved to optimality within the prescribed time limit. However, the additional benefit of stochastic modeling was negligible because of the homothetic structure of the scenarios considered, which preserves the spatial distribution of loads. The results show that spatially heterogeneous scenarios are important for the representation of uncertainty to meaningfully affect reconfiguration decisions.
\end{abstract}

\keywords{reconfiguration, stochastic programming, demand uncertainty, radial distribution networks, technical losses, scenario-based optimization.}

\section{Introduction}

Electricity demand in distribution systems (DSs) is affected by several factors beyond conventional consumption patterns, including the rapid growth of distributed micro- and minigeneration (MMGD, its Portuguese acronym) and electric vehicles. In Brazil, MMGD reached 45.0 gigawatts (GW) of installed capacity and benefited more than 7.2 million consumers by the end of 2025, driven primarily by solar photovoltaic generation \cite{epe-2026}. This new landscape increases demand variability, making efficient and robust operation of distribution networks (DNs) necessary.

This demand variability makes network reconfiguration a relevant problem in power system planning and operation. One of its traditional objectives is to reduce technical losses by changing the status of switching devices while maintaining load supply and the radial structure of the network \cite{civanlar1988,baran1989}.

Distribution network reconfiguration therefore consists of selecting an operating topology by opening and closing the network's sectionalizing or switching devices (hereafter referred to as \textit{switches}). In its deterministic form, the problem assumes known demand values. However, this assumption may produce configurations that are overly tailored to a single operating point. In stochastic distribution network reconfiguration, demand is represented by a set of scenarios with associated probabilities, allowing the same topology decision to be evaluated under different loading conditions.

The scientific literature addresses this uncertainty through different strategies, including scenario-based stochastic programming, robust optimization, and hybrid formulations, combined with linear, conic, or nonlinear power flow models \cite{mahdavi2023}. Computational approaches include exact mixed-integer programming solvers, decomposition techniques, and, in some studies, metaheuristics.

This paper adopts a two-stage stochastic formulation \cite{birge2011}. Topology decisions are made in the first stage and remain common to all scenarios, whereas power flows and other operational variables are determined in the second stage. Uncertainty is initially restricted to demand.

The deterministic equivalent is formulated as a multiscenario mixed-integer second-order cone programming (MISOCP) problem \cite{vielma2017}, incorporating a LinDistFlow approximation \cite{baran1989b,baran1989}, a convex quadratic relation that approximately models losses, explicit radiality constraints, and an optional switching penalty term. Gurobi (\url{https://www.gurobi.com}) was used to solve the model directly. This structure provides a controlled basis for assessing how uncertainty can influence reconfiguration when it is explicitly accounted for.

\section{Literature Review}
\label{sec:revisao_bibliografica}
Technical losses arise from the electrical resistance of lines and equipment throughout the power network and are therefore continuously present, reducing the amount of electricity available from the network \cite{antmann2009,aneel2025}. Distribution networks contain switches (sectionalizing and switching devices) that can be either open or closed. Distribution network reconfiguration (DNR) consists of opening and closing these switches to select a radial topology that supplies the loads and satisfies the network's operating constraints, with the aim of reducing technical losses or other operating costs.

The engineers Merlin and Back \cite{Merlin1975} were the first to identify the possibility of reducing losses through switching operations. Their work was an important milestone, motivating subsequent research and implementations in the field.

Heuristic approaches dominate the literature on DNR. Typically, a feasible spanning tree is used to generate new feasible solutions that improve the objective function \cite{civanlar1988,baran1989,taylor1990,goswami1992,chen1993,cavellucci1997,lyra2003}. Other heuristics adopt a two-step procedure in which an optimistic solution guides radialization strategies toward a near-optimal solution \cite{shirmohammadi1989,wagner1991}. Various metaheuristics, including genetic algorithms and tabu search, have been applied to DNR with satisfactory results \cite{Jeon2000,bueno2004,Subburaj2006,Enacheanu2008,queiroz2009,torres2010,Jakus2020,mahendran2021,otuo_acheampong2023}.

The studies by Civanlar et al.~\cite{civanlar1988} and Baran and Wu~\cite{baran1989} are early milestones in deterministic DNR for loss reduction. Civanlar et al.~\cite{civanlar1988} show that topology changes achieved through switching operations can reduce losses in distribution feeders. Their study makes explicit the relationship between the choice of switches to open or close and the resulting electrical performance, establishing reconfiguration as a combinatorial operating decision.

Baran and Wu~\cite{baran1989} extend this framework by addressing reconfiguration for loss reduction and load balancing. Their work established a radial representation of the problem that became a recurring reference in subsequent studies. In both cases, demand is assumed to be known when the decision is made. The resulting topology is thus optimized for a specified operating point.

This deterministic assumption becomes a limitation when the configuration must remain in operation under different loading levels. A topology that performs well under nominal demand may incur greater losses or operating violations when the load departs from the value used in optimization. Incorporating uncertainty therefore requires a distinction between the topology decision made in advance and the electrical behavior of the network after demand is realized.

In addition to heuristic approaches, classical nonlinear programming methods are also used to formulate and solve DNR \cite{jabr2012,taylor2012,mahdavi2019,mahdavi2021,belotti2025,ayazi2026}.

The concept of \textit{stochastic distribution network reconfiguration} (SDNR) has evolved over recent decades through the convergence of two research streams in electrical engineering: classical network reconfiguration methods and probabilistic/stochastic power flow approaches. Stochastic formulations gained prominence when researchers began combining network reconfiguration with probabilistic mathematical models, such as Monte Carlo techniques and point estimation methods, to represent the inherent uncertainty associated with demand and distributed generation (solar and wind) \cite{kagan1998,Malekpour2013}.

In 2015, Lee et al.~\cite{lee2015} presented a two-stage robust formulation. The radial topology is decided before demand is realized, while a conic optimal power flow problem provides operational recourse. Load variations are represented by a polyhedral set with an uncertainty budget, and the problem is solved by a column-and-constraint generation algorithm that alternates between master and adversarial problems formulated as MISOCPs. On networks with 16 to 94 buses, the robust configurations approximately preserve nominal performance and remain feasible for demands within the uncertainty set considered. The main contribution of this study therefore lies in robust optimization rather than probability-based stochastic programming. Its limitations include the absence of independent AC validation, out-of-sample evaluation, and statistical calibration of the uncertainty set.

In 2020, Akrami, Doostizadeh, and Aminifar~\cite{akrami2020} advanced a data-driven approach combining stochastic and robust elements. Measurements from $\mu$PMUs are organized through fuzzy granulation, producing probability-weighted clusters of net demand. Within each granule, interval uncertainty sets with cardinality constraints protect the solution against adverse realizations. The problem is formulated as a branch-flow-based MISOCP, including radiality, switching costs, and a limit on switching operations, and is solved by column-and-constraint generation. On the 33- and 135-bus systems, the approach reduces the conservatism associated with a single robust uncertainty set and avoids some of the infeasibilities observed in deterministic solutions. However, solution quality depends on how the clusters are constructed and calibrated; limitations also remain concerning the synthetic nature of the data, the independence of the out-of-sample evaluation, and the absence of independent AC validation.

In 2023, Mahdavi, Schmitt, and Jurado~\cite{mahdavi2023} proposed a MISOCP formulation for daily reconfiguration considering variations and uncertainties in load and distributed generation. The model combines binary topology decisions, an SOCP relaxation of the branch flow model, and linear approximations of density functions associated with uncertain deviations. Tests on 33-, 70-, and 118-bus networks indicate lower losses under adverse conditions and shorter solution times than the comparison methods. The study extends the probabilistic representation of uncertainty, but its formulation is monolithic and does not explicitly describe the classical scenario-specific recourse structure. Nonanticipativity constraints, traditional stochastic programming metrics, and out-of-sample evaluations are also not reported. For this reason, its contribution is classified here primarily as the treatment of uncertainty within a MISOCP formulation, rather than as a canonical two-stage stochastic programming formulation.

In 2025, Esmaeilnezhad et al.~\cite{esmaeilnezhad2025} considered dynamic reconfiguration under simultaneous uncertainty in load, wind generation, and photovoltaic generation. Gaussian, Weibull, and Beta distributions are discretized into 45 hourly scenarios, and expected costs are evaluated using AC power flow. The problem incorporates losses, voltage deviation, purchased energy, renewable generation, and switching operations, and is solved using the Coati Optimization Algorithm. On the 33- and 83-bus systems, the dynamic strategy reduces costs and losses relative to the initial and static configurations. The study illustrates the trend toward combining multiple sources of uncertainty with temporal decisions; however, it provides no optimality guarantee, out-of-sample evaluation, or rigorous radiality formulation, making it difficult to separate the effect of stochastic modeling from that of the metaheuristic employed.

More recently, in 2026, Yahia and Gheith~\cite{yahia2026} formulated a stochastic MINLP based on sample average approximation for the dynamic reconfiguration of a meshed network containing three microgrids and 15 households. Demand and renewable generation scenarios are constructed empirically from approximately two years of data and used in an hourly schedule that maximizes the minimum expected load factor. N--1 and N--2 contingency tests indicate full demand satisfaction and greater stability for the configuration with three microgrids, although a single unit exhibits higher utilization. The use of empirical data and sample average approximation brings the formulation closer to a classical stochastic framework; however, the model does not represent voltages or reactive power, does not account for switching operations, and does not provide a rigorous connectivity formulation, out-of-sample validation, or sufficient computational details for reproducibility.

The chronological sequence of the studies considered reveals a progressive change in the treatment of uncertainty. The deterministic literature initially establishes DNR as a problem of selecting a radial topology to reduce losses. From the early 2010s (between 2011 and 2015), robust formulations with first- and second-stage decisions began to appear; subsequently, operational data, probabilities, parametric distributions, and scenario generation were incorporated into the decision-making process. More recent studies also extend the temporal dimension of the problem and combine load, renewable generation, and microgrids within a single formulation.

This evolution, however, makes it difficult to isolate the benefit of accounting for demand uncertainty. In several studies, new sources of uncertainty are introduced together with distributed generation, multiperiod operation, new objectives, or metaheuristics. Moreover, certain limitations recur throughout the reviewed literature: absence of out-of-sample evaluation, lack of independent electrical validation, insufficiently explicit radiality or connectivity formulations, absence of optimality guarantees, and limited computational documentation. The studies reviewed also lack a uniform protocol for comparing a deterministic solution with its stochastic counterpart while holding the network representation, objective, and other modeling assumptions constant.

The present study addresses this issue. Rather than simultaneously combining multiple sources of uncertainty and technologies, it initially considers only load uncertainty. The topology is a first-stage decision common to all scenarios, while the operational variables are adjusted in the second stage.

This scope allows the effect of stochastic modeling on the selected configuration to be assessed directly and enables comparisons of expected losses, worst-case scenario losses, and electrical violations within the same mathematical framework. The intended contribution is therefore not to increase the number of uncertain sources, but to establish a controlled and reproducible analysis of the value of explicitly accounting for demand uncertainty in distribution network reconfiguration.

\section{Fundamental Concepts}
\label{sec:conceitos_fundamentais}

The review in the previous section shows that distribution network reconfiguration has evolved from a deterministic topology selection problem to formulations that explicitly incorporate operational uncertainty. Understanding the scope adopted in this study requires distinguishing the topological elements of the network, the reconfiguration decisions, the electrical approximation employed, and the way demand uncertainty changes the structure of the problem. This section introduces these concepts before presenting the complete mathematical formulation.

Methodologically, this is a quantitative study based on mathematical modeling and computational experimentation. The strategy is incremental: it starts from a validated deterministic representation of reconfiguration and introduces demand uncertainty through scenarios while retaining the same candidate network and primary electrical criterion. This choice allows approaches to be compared under controlled conditions and separates the effect of the topology decision from that of load variation.

\subsection{Graph-theoretic representation of the distribution network}
\label{subsec:grafos}

A distribution network is represented by a graph
\begin{equation}
G=(\mathcal{N},\mathcal{E}),
\end{equation}
where $\mathcal{N}$ is the set of buses and $\mathcal{E}$ is the set of physical branches available for the configuration. Each branch $e=\{i,j\}\in\mathcal{E}$ represents an electrical connection between buses $i$ and $j$ and has electrical parameters and an initial operating status.

In the implementation, the physical graph is initially treated as undirected for topological purposes. A root corresponding to the substation is then used; when multiple substations exist, preprocessing may introduce an artificial root connected to them by zero-impedance branches. The candidate branches are then oriented from this root according to levels obtained by breadth-first search. This orientation defines the predecessor and successor relationships used in the formulation, while the flows $P$ and $Q$ remain signed variables.

Several graph-theoretic concepts are particularly important:
\begin{itemize}
    \item \textbf{Path:} a sequence of nodes connected by consecutive branches.
    \item \textbf{Cycle:} a closed path that returns to its starting node without repeating any other node. A radial topology cannot contain cycles in the subgraph of closed branches.
    \item \textbf{Connected graph:} a graph in which a path exists between any pair of nodes.
    \item \textbf{Tree:} a connected acyclic graph. A tree with $|\mathcal{N}|$ nodes has $|\mathcal{N}|-1$ branches.
    \item \textbf{Rooted tree:} a tree oriented away from a root, in which every nonroot node has exactly one predecessor.
    \item \textbf{Forest:} a collection of disjoint trees. An artificial root allows networks with multiple substations to be represented by a single rooted structure during optimization.
    \item \textbf{Support graph of a solution:} the subgraph formed by branches whose binary status variable equals one.
    \item \textbf{Set of cycles $\mathcal{C}$:} cycles identified in the support graph during the solver's search and used to generate cycle elimination inequalities.
\end{itemize}

Radiality is enforced through a combination of three mechanisms. First, every nonroot bus must have exactly one selected incoming branch. Second, the total number of closed branches is fixed at $|\mathcal{N}|-1$. Third, any cycles present in candidate integer solutions are eliminated dynamically. Together, these conditions characterize the rooted tree structure used in the model. The resulting topology is checked again during the subsequent electrical evaluation.

\subsection{Concepts related to deterministic reconfiguration}
\label{subsec:conceitos_deterministicos}

In deterministic reconfiguration, active and reactive power demands at the buses are assumed to be known. The initial configuration is defined by the branch statuses, distinguishing initially closed from initially open connections. The reconfiguration decision consists of changing some of these statuses to obtain a new radial topology with improved operational performance.

A \textbf{switching operation} is a change in the status of a branch counted as switchable: opening an initially closed connection or closing an initially open connection. For each branch $(i,j)$ with initial status $y_{ij}^{0}$ and final status $y_{ij}$, this change can be represented by $|y_{ij}-y_{ij}^{0}|$. The general formulation in this study allows these changes to be penalized in the objective function. However, this penalty is optional and can be disabled, in which case the number of switching operations is used only as a metric of the resulting solution.

A \textbf{feasible configuration} must simultaneously satisfy topological and electrical conditions. Topologically, the buses must form a rooted radial structure. Electrically, the configuration must admit power balance and voltages within the bounds used in optimization.

Topology selection uses a convex approximation of distribution power flow. The model employs power balances without explicit loss terms, LinDistFlow relations for voltage drops, and a convex quadratic inequality relating $P$, $Q$, and an auxiliary variable $\ell$. This structure yields a MISOCP. After optimization, the topology is evaluated separately using an approximate classical Baran--Wu radial power flow method to compute approximate losses, voltage profiles, and violations. The solution of the optimization problem is thus distinguished from the \textit{ex post} electrical evaluation, without treating the latter as an exact AC power flow calculation.

\subsection{Concepts related to stochastic reconfiguration}
\label{subsec:conceitos_estocasticos}

In stochastic reconfiguration, load is no longer represented by a single deterministic vector. Let $\boldsymbol{\xi}$ denote the random variable representing network demand. In this study, its distribution is approximated by a finite set of scenarios
\begin{equation}
\Omega=\{1,\ldots,S\},
\end{equation}
where each scenario $s\in\Omega$ has probability $\pi_s$, with
\begin{equation}
\pi_s\geq 0,
\qquad
\sum_{s\in\Omega}\pi_s=1.
\end{equation}

The two-stage structure follows from the order in which information becomes available. Switching decisions are made before the demand scenario is known and constitute \textit{first-stage} decisions. A single topology must therefore be shared by all scenarios.

After demand is realized, power flows, voltages, and the auxiliary loss variable take scenario-specific values. These quantities constitute \textit{second-stage} or \textit{recourse} decisions. Nonanticipativity is implemented structurally through a single binary topology variable for each branch, with no scenario index, while the operational variables are replicated for each $s\in\Omega$.

The basic stochastic criterion is the expected value. If $\widehat L_s$ represents the approximate electrical criterion for scenario $s$, then
\begin{equation}
\mathbb{E}[\widehat L]
=
\sum_{s\in\Omega}\pi_s\widehat L_s.
\end{equation}
The experimental evaluation may also consider the maximum loss across scenarios, minimum voltages, electrical violations, and stochastic programming metrics such as the value of the stochastic solution, without changing the first- and second-stage structure.

\subsection{Illustrative example}
\label{subsec:exemplo_didatico}

Consider a simplified network with four buses,
\begin{equation}
\mathcal{N}=\{0,1,2,3\},
\end{equation}
where bus $0$ represents the substation. Suppose that there are four candidate branches,
\begin{equation}
\mathcal{E}
=
\{\{0,1\},\{1,2\},\{2,3\},\{0,3\}\}.
\end{equation}
The initial configuration keeps branches $\{0,1\}$, $\{1,2\}$, and $\{2,3\}$ closed and the tie branch $\{0,3\}$ open. The active network is therefore the path
\begin{equation}
0-1-2-3,
\end{equation}
which forms a tree.

If branch $\{0,3\}$ is closed without any other operation, the following cycle is formed:
\begin{equation}
0-1-2-3-0.
\end{equation}
To restore radiality, at least one of the other branches in the cycle must be opened. For example, closing $\{0,3\}$ and opening $\{1,2\}$ yields the active branches
\begin{equation}
\{\{0,1\},\{0,3\},\{2,3\}\},
\end{equation}
which again form a tree. This change corresponds to two switching operations.

In the deterministic case, the two topologies can be compared under a single loading condition. In the stochastic case, suppose, for illustration only, that there are three scenarios with low, nominal, and high loads, with probabilities $0.25$, $0.50$, and $0.25$, respectively. The first-stage decision must select \emph{a single} topology before the scenario is known. For each candidate topology, operational variables are determined in all scenarios, and the corresponding electrical criteria are weighted by their probabilities.

This example highlights the central conceptual difference between the two approaches. Deterministic reconfiguration seeks the best topology for a known condition; stochastic reconfiguration seeks a common topology whose performance is evaluated over a distribution of possible operating conditions.

\subsection{Description of the problem addressed}
\label{subsec:descricao_problema}

The problem studied considers a distribution network with a known candidate topology, branch electrical parameters, an initial configuration, and demand values associated with the buses. In the stochastic case, demand is specified by a finite set of scenarios with known probabilities.

The central decision is to select branch statuses so that the resulting topology is radial. The same topology must be used in all scenarios because switching decisions are made before demand is realized. For each scenario, the electrical operation of the selected topology is determined by scenario-specific continuous variables.

The primary performance criterion is the reduction of expected approximate losses. The general formulation also includes a penalty term for changes relative to the initial configuration. This term allows the trade-off between electrical performance and operational effort to be studied; when its weight is zero, the topology is selected solely according to the electrical criterion, and the number of switching operations is calculated a posteriori.

\subsubsection{Loss calculation and approximation}
\label{subsubsec:calculo_perdas}

The physical active power loss in a closed branch $(i,j)$ is, in principle,
\begin{equation}
P^{\mathrm{loss}}_{ij}=r_{ij}|I_{ij}|^2,
\label{eq:perda_fisica_ramo}
\end{equation}
where $r_{ij}$ is the branch resistance and $I_{ij}$ is the branch current. In a full branch flow formulation, the relationship between current, power, and voltage can be written as
\begin{equation}
P_{ij}^2+Q_{ij}^2 = v_i\ell_{ij},
\end{equation}
with $v_i=|V_i|^2$ and $\ell_{ij}=|I_{ij}|^2$.

The formulation implemented in this study uses a simpler approximation. In the per-unit (pu) system, assuming a voltage close to $1$ pu, the nonnegative variable $\ell_{ij}^{s}$ is introduced and the following inequality is imposed:
\begin{equation}
\left(P_{ij}^{s}\right)^2+
\left(Q_{ij}^{s}\right)^2
\leq
\ell_{ij}^{s}.
\label{eq:loss_conic_concept}
\end{equation}
Since $\ell_{ij}^{s}$ appears with a positive coefficient in the objective function, minimization tends to make the inequality tight. The scenario loss criterion is then
\begin{equation}
\widehat L_s
=
\sum_{(i,j)\in\mathcal{A}}
r_{ij}\ell_{ij}^{s}.
\label{eq:loss_proxy_concept}
\end{equation}
This expression approximates Joule losses while preserving conic convexity; it should not be interpreted as an exact AC power flow model.

After optimization, the topology is fixed and evaluated scenario by scenario using a routine based on the approximate classical Baran--Wu DistFlow model. For each branch, the evaluation uses downstream flows, computes the approximate voltage drop, and reports losses as
\begin{equation}
\widetilde P^{\mathrm{eval}}_{\mathrm{loss},s}
=
S_{\mathrm{base}}
\sum_{(i,j)\in\mathcal{E}^{\mathrm{on}}}
r_{ij}^{\mathrm{pu}}
\left[
\left(P_{ij,s}^{\mathrm{pu}}\right)^2+
\left(Q_{ij,s}^{\mathrm{pu}}\right)^2
\right].
\label{eq:perda_avaliacao_cenario}
\end{equation}
The superscript ``eval'' emphasizes that this metric is produced by the approximate evaluator. The corresponding expected loss is
\begin{equation}
\mathbb{E}\!\left[\widetilde P^{\mathrm{eval}}_{\mathrm{loss}}\right]
=
\sum_{s\in\Omega}\pi_s
\widetilde P^{\mathrm{eval}}_{\mathrm{loss},s},
\label{eq:perda_esperada_eval}
\end{equation}
and the maximum loss across scenarios is
\begin{equation}
\widetilde P^{\max}_{\mathrm{loss}}
=
\max_{s\in\Omega}
\widetilde P^{\mathrm{eval}}_{\mathrm{loss},s}.
\label{eq:perda_max_eval}
\end{equation}

When comparing a reference configuration with an optimized configuration, the percentage loss reduction, using the same evaluation metric, is
\begin{equation}
\Delta P_{\mathrm{loss}}(\%)
=
100
\frac{
P^{\mathrm{ref}}_{\mathrm{loss}}-P^{\mathrm{opt}}_{\mathrm{loss}}
}{P^{\mathrm{ref}}_{\mathrm{loss}}}.
\label{eq:reducao_percentual_perdas}
\end{equation}

This separation between model and evaluation is important. Optimization uses an approximate conic representation to select the topology; the subsequent evaluation uses an approximate radial power flow calculation with the topology fixed. Both levels are consistent with the classical approximation adopted, but neither is presented as a substitute for high-fidelity nonlinear AC validation.

In summary, the problem addressed consists of choosing, before demand is realized, a single radial configuration with good expected electrical performance and, when desired, a low reconfiguration cost. The corresponding mathematical formulation is presented in the next section.

\section{Mathematical Model}
\label{sec:modelo_matematico}

This section brings together the concepts presented in the previous section in a formulation for stochastic distribution network reconfiguration under demand uncertainty. The topology is selected before demand is realized and remains common to all scenarios, whereas flows, voltages, and auxiliary variables are determined separately for each scenario. The deterministic equivalent formulation contains binary variables and convex quadratic constraints and is therefore classified as a MISOCP problem; in the Gurobi interface, these relations are entered as convex quadratic constraints in a MIQCP.

\subsection{Optimization problem formulation}
\label{subsec:formulacao_otimizacao}

After preprocessing, consider the candidate directed graph $G=(\mathcal{N},\mathcal{A})$, with root $r\in\mathcal{N}$. For a bus $i$, let $\delta^-(i)$ and $\delta^+(i)$ denote the sets of directed branches entering and leaving $i$, respectively. The orientation is used to structure the tree and write the balance equations; the flows $P$ and $Q$ may be positive or negative.

The following sets are used:
\begin{itemize}
    \item $\mathcal{N}$: set of buses, including the root;
    \item $\mathcal{A}$: set of candidate directed branches;
    \item $\mathcal{E}^{sw}\subseteq\mathcal{A}$: branches whose switching operations can be counted and penalized;
    \item $\Omega$: finite set of demand scenarios;
    \item $\mathcal{C}$: cycles identified in the support graph during optimization.
\end{itemize}

For each bus $i$, branch $(i,j)$, and scenario $s$, the following parameters are considered:
\begin{itemize}
    \item $p_i^s$ and $q_i^s$: active and reactive power demands in pu;
    \item $\pi_s$: scenario probability, with $\sum_{s\in\Omega}\pi_s=1$;
    \item $r_{ij}$ and $x_{ij}$: branch resistance and reactance in pu;
    \item $\underline V_i$ and $\overline V_i$: voltage magnitude bounds;
    \item $V_r$: root voltage magnitude;
    \item $M_{ij,s}^{P}$, $M_{ij,s}^{Q}$, and $M_{ij,s}^{\ell}$: bounds used to couple flows with branch status;
    \item $M_{ij,s}^{v}$: constant used to relax the voltage drop equation when the branch is open;
    \item $y_{ij}^{0}\in\{0,1\}$: initial branch status;
    \item $c_{ij}^{sw}>0$: weight associated with a status change;
    \item $\gamma\geq0$: overall weight of the switching penalty.
\end{itemize}

The first-stage variables are
\begin{align}
y_{ij}&\in\{0,1\},
&&\forall(i,j)\in\mathcal{A},
\label{eq:y_domain_v07}\\
u_{ij}&\geq0,
&&\forall(i,j)\in\mathcal{E}^{sw},
\label{eq:u_domain_v07}
\end{align}
where $y_{ij}=1$ indicates a closed branch and $u_{ij}$ represents the status change when the penalty is enabled.

For each scenario $s\in\Omega$, the second-stage variables are
\begin{align}
P_{ij}^{s},Q_{ij}^{s}&\in\mathbb{R},
&&\forall(i,j)\in\mathcal{A},
\label{eq:PQ_domain_v07}\\
\ell_{ij}^{s}&\geq0,
&&\forall(i,j)\in\mathcal{A},
\label{eq:l_domain_v07}\\
v_i^s&\geq0,
&&\forall i\in\mathcal{N}.
\label{eq:v_domain_v07}
\end{align}
The variable $v_i^s$ represents the squared voltage magnitude, while $\ell_{ij}^s$ is the auxiliary variable used in the convex approximation of losses.

\subsubsection{Objective function}

For each scenario, define
\begin{equation}
\widehat L_s
=
\sum_{(i,j)\in\mathcal{A}}
r_{ij}\ell_{ij}^{s}.
\label{eq:loss_proxy_s_v07}
\end{equation}
The general formulation minimizes the expected electrical criterion plus an optional switching penalty:
\begin{equation}
\min\;
\sum_{s\in\Omega}\pi_s\widehat L_s
+
\gamma
\sum_{(i,j)\in\mathcal{E}^{sw}}
c_{ij}^{sw}u_{ij}.
\label{eq:stochastic_objective_v07}
\end{equation}
The first term selects a topology by considering all scenarios simultaneously. The second allows switching effort to be controlled. In the experiments presented in this paper, $\gamma=0$; thus, the penalty remains part of the general formulation but does not affect the results currently reported.

\subsubsection{Topological constraints and radiality}

Each nonroot bus must have exactly one selected predecessor:
\begin{equation}
\sum_{(h,i)\in\delta^-(i)}y_{hi}=1,
\qquad
\forall i\in\mathcal{N}\setminus\{r\}.
\label{eq:indegree_v07}
\end{equation}
The cardinality of the topology is enforced by
\begin{equation}
\sum_{(i,j)\in\mathcal{A}}y_{ij}
=
|\mathcal{N}|-1.
\label{eq:tree_cardinality_v07}
\end{equation}

In addition to these constraints, for every cycle $C$ identified in the support graph of a candidate solution, the following inequality is added:
\begin{equation}
\sum_{(i,j)\in C}y_{ij}
\leq
|C|-1.
\label{eq:cycle_elimination_v07}
\end{equation}

Equation~\eqref{eq:cycle_elimination_v07}
prevents all branches of a cycle from remaining closed. Together with the unique-predecessor and cardinality constraints, the cycle inequalities ensure the rooted radial structure used in the implementation.

\subsubsection{Power balance by scenario}

For each nonroot bus and scenario $s$, the balances are written without explicit loss terms:
\begin{equation}
\sum_{(h,i)\in\delta^-(i)}P_{hi}^{s}
-
\sum_{(i,j)\in\delta^+(i)}P_{ij}^{s}
=
p_i^s,
\qquad
\forall i\neq r,\;s\in\Omega,
\label{eq:active_balance_v07}
\end{equation}
\begin{equation}
\sum_{(h,i)\in\delta^-(i)}Q_{hi}^{s}
-
\sum_{(i,j)\in\delta^+(i)}Q_{ij}^{s}
=
q_i^s,
\qquad
\forall i\neq r,\;s\in\Omega.
\label{eq:reactive_balance_v07}
\end{equation}
The root represents the network's power source; its injection is implicitly determined by the flows leaving the root.

\subsubsection{Coupling topology, flows, and losses}

Flows and the auxiliary variable are set to zero when a branch is open:
\begin{align}
-M_{ij,s}^{P}y_{ij}
\leq P_{ij}^{s}
&\leq M_{ij,s}^{P}y_{ij},
\label{eq:P_activation_v07}\\
-M_{ij,s}^{Q}y_{ij}
\leq Q_{ij}^{s}
&\leq M_{ij,s}^{Q}y_{ij},
\label{eq:Q_activation_v07}\\
0\leq\ell_{ij}^{s}
&\leq M_{ij,s}^{\ell}y_{ij},
\label{eq:l_activation_v07}
\end{align}
for all $(i,j)\in\mathcal{A}$ and $s\in\Omega$.

The convex approximation of the flow magnitude is
\begin{equation}
\left(P_{ij}^{s}\right)^2+
\left(Q_{ij}^{s}\right)^2
\leq
\ell_{ij}^{s},
\qquad
\forall(i,j)\in\mathcal{A},\;s\in\Omega.
\label{eq:current_cone_v07}
\end{equation}
Equation~\eqref{eq:current_cone_v07} is convex and second-order cone representable. Since $r_{ij}\ell_{ij}^{s}$ has a nonnegative coefficient in the objective function, the solution tends to use the smallest feasible value of $\ell_{ij}^{s}$.

\subsubsection{Voltage drops and bounds}

The root voltage is fixed by
\begin{equation}
v_r^s=V_r^2,
\qquad
\forall s\in\Omega.
\label{eq:root_voltage_v07}
\end{equation}
For the remaining buses,
\begin{equation}
\underline V_i^2
\leq
v_i^s
\leq
\overline V_i^2,
\qquad
\forall i\in\mathcal{N},\;s\in\Omega.
\label{eq:voltage_limits_v07}
\end{equation}

When the branch is closed, the LinDistFlow approximation is used:
\begin{equation}
v_j^s
=
v_i^s-2\left(r_{ij}P_{ij}^{s}+x_{ij}Q_{ij}^{s}\right).
\label{eq:lindistflow_ideal_v07}
\end{equation}
Coupling with the binary variable is achieved through
\begin{align}
v_j^s
&\leq
v_i^s-2\left(r_{ij}P_{ij}^{s}+x_{ij}Q_{ij}^{s}\right)
+M_{ij,s}^{v}(1-y_{ij}),
\label{eq:voltage_drop_upper_v07}\\
v_j^s
&\geq
v_i^s-2\left(r_{ij}P_{ij}^{s}+x_{ij}Q_{ij}^{s}\right)
-M_{ij,s}^{v}(1-y_{ij}).
\label{eq:voltage_drop_lower_v07}
\end{align}
Thus, the voltage drop relation is enforced when $y_{ij}=1$ and relaxed when $y_{ij}=0$.

\subsubsection{Switching penalty}

For branches whose switching operations are counted,
\begin{align}
u_{ij}&\geq y_{ij}-y_{ij}^{0},
\label{eq:switching_1_v07}\\
u_{ij}&\geq y_{ij}^{0}-y_{ij},
\label{eq:switching_2_v07}
\end{align}
for all $(i,j)\in\mathcal{E}^{sw}$. When $\gamma c_{ij}^{sw}>0$, minimization implies
\begin{equation}
u_{ij}=|y_{ij}-y_{ij}^{0}|.
\end{equation}
With unit weights, the switching term represents the number of status changes; with different weights, it represents a weighted operating cost. When $\gamma=0$, as in the current experiments, the auxiliary penalty variables are not required for topology selection, and the number of switching operations is computed during postprocessing by comparing $y_{ij}$ with $y_{ij}^{0}$.

\subsubsection{Compact form}

The deterministic equivalent can be summarized as
\begin{equation}
\begin{aligned}
\min\quad
&
\sum_{s\in\Omega}\pi_s
\sum_{(i,j)\in\mathcal{A}}r_{ij}\ell_{ij}^{s}
+
\gamma\sum_{(i,j)\in\mathcal{E}^{sw}}c_{ij}^{sw}u_{ij}
\\[1mm]
\text{subject to}\quad
&
\eqref{eq:indegree_v07}-\eqref{eq:cycle_elimination_v07},\\
&
\eqref{eq:active_balance_v07}-\eqref{eq:reactive_balance_v07},\\
&
\eqref{eq:P_activation_v07}-\eqref{eq:current_cone_v07},\\
&
\eqref{eq:root_voltage_v07}-\eqref{eq:voltage_drop_lower_v07},\\
&
\eqref{eq:switching_1_v07}-\eqref{eq:switching_2_v07},\\
&
y_{ij}\in\{0,1\},\quad u_{ij}\geq0,\\
&P_{ij}^{s},Q_{ij}^{s}\in\mathbb{R},\quad
\ell_{ij}^{s},v_i^s\geq0.
\end{aligned}
\label{eq:compact_model_v07}
\end{equation}

The only discrete variables are the branch statuses, and nonlinearity is restricted to the convex inequalities in Equation~\eqref{eq:current_cone_v07}. For a finite set of scenarios, the formulation is therefore a multiscenario MISOCP. The deterministic case is recovered by letting $\Omega$ contain a single scenario.

\subsection{Model description}
\label{subsec:descricao_modelo}

In the first stage, the model selects the values of $y_{ij}$ and, when the penalty is used, the variables $u_{ij}$ associated with status changes. These decisions are common to all scenarios. Radiality is enforced by the predecessor, cardinality, and cycle elimination constraints.

In the second stage, each scenario has its own variables $P_{ij}^{s}$, $Q_{ij}^{s}$, $\ell_{ij}^{s}$, and $v_i^s$. The absence of a scenario index in $y_{ij}$ prevents a different configuration from being selected after observing demand and implements the nonanticipativity condition.

The objective function accounts for probability-weighted electrical performance and allows a switching cost term. The general formulation therefore supports the study of the trade-off between expected approximate losses and operational effort. In the experimental setting used in this paper, $\gamma=0$, so switching operations do not influence the optimal decision and are reported as an ex post metric.

The electrical approximation combines lossless balances, LinDistFlow, and the convex relation $P^2+Q^2\leq\ell$. This choice keeps the problem within the MISOCP class and allows the solver to certify solution quality. The resulting topology is subsequently subjected to an approximate Baran--Wu radial evaluation in each scenario to produce the electrical metrics presented in the results.

\subsection{Limitations and extensions}
\label{subsec:limitacoes_expansoes}

The formulation was deliberately kept within a controlled scope to isolate the effect of demand uncertainty. Its main limitations are:
\begin{itemize}
    \item the model is single-period and selects a single topology common to all scenarios;
    \item only demand is treated as uncertain, and scenario probabilities are assumed to be known;
    \item the power balances do not explicitly include branch losses;
    \item the relation $P^2+Q^2\leq\ell$ approximates current by omitting the sending-end voltage factor multiplying $\ell$, as would be included in a more complete branch flow formulation;
    \item the voltage drop is modeled using LinDistFlow, omitting the quadratic term associated with losses;
    \item the bounds used to activate $P$, $Q$, and $\ell$ serve a modeling purpose and do not, by themselves, replace a detailed representation of equipment thermal limits;
    \item the subsequent electrical evaluation also uses an approximate radial power flow method rather than high-fidelity nonlinear AC power flow;
    \item the objective is risk-neutral and includes neither measures such as CVaR nor robust uncertainty sets;
    \item although the model supports switching penalties, this mechanism is not enabled in the experiments presented in this version of the paper.
\end{itemize}

Natural extensions include a branch flow MISOCP formulation with full current--voltage coupling, independent AC validation, distributed generation and storage, spatially heterogeneous scenarios, multiperiod models, systematic analysis of parameter $\gamma$, risk measures, and decomposition techniques for larger scenario sets. These extensions are not necessary for the controlled analysis developed in this paper, but they establish directions for future work.

\section{Solution Techniques}
\label{sec:tecnicas_solucao}

The formulation in Section~\ref{sec:modelo_matematico} is a two-stage stochastic program represented by its multiscenario MISOCP deterministic equivalent. The binary topology variables are shared by all scenarios, whereas the continuous operational variables are replicated for each demand realization. The adopted strategy consists of solving this deterministic equivalent monolithically with Gurobi and handling cycles through \textit{lazy constraints} added during the search.

\subsection{Techniques used in the literature}
\label{subsec:tecnicas_literatura}

The reviewed studies employ different strategies depending on the representation of uncertainty and the electrical structure. Lee et al.~\cite{lee2015} and Akrami et al.~\cite{akrami2020} use column-and-constraint generation to decompose two-stage robust problems with adversarial subproblems. Mahdavi, Schmitt, and Jurado~\cite{mahdavi2023} solve a monolithic MISOCP formulation. Esmaeilnezhad et al.~\cite{esmaeilnezhad2025} use a metaheuristic with scenario-specific electrical evaluation, whereas Yahia and Gheith~\cite{yahia2026} employ sample average approximation in a stochastic MINLP.

These studies highlight two recurring challenges: adequately representing uncertainty and controlling the combinatorial complexity of the topology. In this paper, the scenario set is deliberately moderate in size, and the deterministic equivalent remains convex in the continuous variables. For this reason, a monolithic solution approach using an exact commercial solver was selected, retaining the ability to record solution status, dual bound, \textit{MIP gap}, and processing time.

\subsection{Model construction and solution with Gurobi}
\label{subsec:gurobi}

The implementation is developed in Python using the \texttt{gurobipy} interface. For each instance, the program reads and validates the network, constructs and orients the graph from the root, loads the scenarios, creates the deterministic equivalent, configures the solver, and starts optimization. The variables $y_{ij}$ are common to all scenarios; the variables $P_{ij}^{s}$, $Q_{ij}^{s}$, $\ell_{ij}^{s}$, and $v_i^s$ are created for each $s\in\Omega$. Nonanticipativity is therefore implemented structurally.

The relation
\begin{equation}
\left(P_{ij}^{s}\right)^2+\left(Q_{ij}^{s}\right)^2\leq\ell_{ij}^{s}
\end{equation}
is added to Gurobi as a convex quadratic constraint. The solver handles the binary variables within its \textit{branch-and-bound}/\textit{branch-and-cut} framework, combined with internal routines for convex quadratic problems.

The reference run used for the current results adopts
\begin{equation}
\texttt{TimeLimit}=600\;\text{s},
\qquad
\texttt{MIPGap}=10^{-4},
\qquad
\texttt{Threads}=4,
\label{eq:gurobi_parameters_v07}
\end{equation}
together with
\begin{equation}
\texttt{Presolve}=1,
\qquad
\texttt{NumericFocus}=2.
\label{eq:gurobi_parameters_extra_v07}
\end{equation}
All parameters not explicitly configured retain their Gurobi default values. For each run, the solver status, objective value, best available bound, \textit{MIP gap}, and computation time are retrieved.

\subsection{Warm start from the initial configuration}
\label{subsec:warm_start}

The run is configured with \texttt{warm\_start: sma\_full}. Before optimization, the implementation checks whether the initially active topology forms a valid directed tree rooted at the designated root. When this condition is satisfied, an initial value is assigned to every binary variable $y_{ij}$ according to the initial network status.

In \texttt{sma\_full} mode, initial values are also constructed for the continuous variables of each scenario. For the \texttt{method1\_classic\_approx\_distflow} formulation, initial flows are obtained by accumulating downstream loads in the initial tree, variable $\ell$ is assigned values consistent with $P^2+Q^2$, and voltages are initialized using the LinDistFlow relation along the tree. These values are provided to Gurobi as a \textit{MIP start}. The procedure changes neither the feasible set nor the optimality criterion; its purpose is to provide the solver with a structured initial solution that may accelerate the discovery of incumbents.

\subsection{Dynamic separation of cycle elimination constraints}
\label{subsec:cycle_separation}

In addition to the predecessor and cardinality constraints included in the initial model, the implementation uses the inequality
\begin{equation}
\sum_{(i,j)\in C}y_{ij}
\leq |C|-1
\label{eq:cycle_cut_solution_v07}
\end{equation}
to eliminate cycles found in candidate integer solutions. Rather than enumerating all simple cycles in the graph in advance, these inequalities are generated on demand.

The \textit{lazy constraint} mechanism is enabled by
\begin{equation}
\texttt{LazyConstraints}=1.
\end{equation}
When Gurobi reaches the \texttt{MIPSOL} event, the \textit{callback} retrieves the incumbent values of $y$, selects the branches with $y_{ij}>0.5$, constructs the support graph, and invokes cycle detection. For each cycle returned, Equation~\eqref{eq:cycle_cut_solution_v07} is added through \texttt{cbLazy}; the search continues without accepting a solution that violates acyclicity as feasible.

The detection routine operates on the undirected representation of the support graph. The code provides two \textit{backends}: an optional native implementation and a Python implementation based on NetworkX. Under the default \texttt{auto} setting, the native \textit{backend} is attempted first; if it is unavailable, the Python version is used. The cut logic is the same in both cases. The current implementation does not maintain a persistent table of signatures for previously separated cycles; therefore, no explicit duplicate-handling mechanism is attributed to the method beyond the handling performed by the search process itself.

\subsection{Solution workflow and ex post evaluation}
\label{subsec:solution_workflow}

The computational procedure for each instance follows these steps:
\begin{enumerate}
    \item read the network and validate the basic data;
    \item construct the candidate graph, define the root, and orient the branches;
    \item load the scenarios and their probabilities;
    \item construct the MISOCP deterministic equivalent with a shared topology;
    \item apply the Gurobi settings and the full \textit{warm start};
    \item enable \textit{lazy constraints} and solve the model with the cycle \textit{callback};
    \item extract the topology, switching operations, status, objective, \textit{gap}, and time;
    \item fix the resulting topology and evaluate it separately in each scenario using the approximate classical Baran--Wu radial power flow method;
    \item compute approximate losses in kW, minimum and maximum voltages, voltage violations, and other evaluation metrics.
\end{enumerate}

In the ex post stage, the evaluator reconstructs a directed graph containing only closed branches. The routine requires a radial structure consistent with the root: the graph must be acyclic, and every nonroot bus must have exactly one predecessor. Flows are then accumulated from downstream loads, and voltages are propagated from the root to the leaves using the LinDistFlow relation. The reported losses are the approximate losses defined in Equation~\eqref{eq:perda_avaliacao_cenario}; for this reason, the results are not presented as exact AC validation.

Switching penalties remain available in the implementation through the \texttt{unit} and \texttt{weighted} modes. However, the current results use \texttt{switching\_penalty\_mode: none} and $\gamma=0$. Consequently, the number of switching operations is computed after optimization by comparing the final topology with the initial status, without affecting the objective function used in these experiments.

\subsection{Computational challenges}
\label{subsec:desafios_solucao}

The first challenge is the growth of the deterministic equivalent with the number of scenarios. The binary topology variables remain shared, but $P$, $Q$, $\ell$, $v$, and their constraints are replicated for each scenario. Thus, the cost of processing each node in the search tree increases with $|\Omega|$ and network size.

The second challenge stems from the relaxations and bounds used to couple the topology with the continuous variables. The implementation computes branch- and scenario-specific bounds from downstream demands and uses these values to activate $P$, $Q$, and $\ell$, and to set the relaxation constants in the voltage equations. Excessively large bounds weaken the relaxation and may increase search effort; overly restrictive values may exclude feasible solutions.

The third challenge concerns cycle separation. Dynamic generation reduces the number of constraints in the initial model but introduces overhead in the \textit{callback}. Efficiency depends on rapidly identifying relevant cycles and on the interaction of these inequalities with the existing predecessor and cardinality constraints.

Finally, advancing beyond the methodological results in the literature requires combining a certificate of solution quality, a rigorous topological representation, explicit treatment of uncertainty, and transparent electrical evaluation. The adopted strategy emphasizes these properties through the monolithic solution of the MISOCP, recording of the \textit{MIP gap}, dynamic generation of cycle cuts, and a separate ex post evaluation. For larger networks or scenario sets, future extensions may explore scenario decomposition, stronger relaxations, and higher-fidelity electrical validation.

\section{Computational Experiments}
\label{sec:experimentos}

This section evaluates the proposed stochastic formulation on a set of networks with different sizes and structures. The analysis uses the final reports from the computational experiments and distinguishes three aspects: the electrical performance of the resulting topology, the effect of the stochastic representation of demand, and the solver's computational performance. All losses presented in this section correspond to the \textit{ex post} evaluation using the approximate classical Baran--Wu radial power flow method, rather than directly to the internal value of the MISOCP objective function.

\subsection{Computational environment and implementation}
\label{subsec:ambiente_experimental}

The experiments were run on an Apple M2 computer with 16~GB of memory, running macOS Tahoe~26.5.1, Python~3.13.11, and Gurobi~12.0.2. The implementation is the same as that described in Section~\ref{sec:tecnicas_solucao}, using the \texttt{method1\_classic\_approx\_distflow} formulation and a monolithic solution of the MISOCP deterministic equivalent.

For all instances, the following settings were used:
\begin{equation}
\texttt{MIPGap}=10^{-4},\qquad
\texttt{TimeLimit}=600\ \mathrm{s},\qquad
\texttt{Threads}=4,
\label{eq:exp_solver_params}
\end{equation}
with \texttt{Presolve}=1, \texttt{NumericFocus}=2, and \texttt{warm\_start}=\texttt{sma\_full}. Cycle elimination constraints were handled as \textit{lazy constraints}. Switching penalties remained available in the model but were disabled in this set of experiments, with \texttt{switching\_penalty\_mode}=\texttt{none} and $\gamma=0$.

In the optimization, the configured voltage bounds were $0.60\leq V\leq1.05$~pu. The \textit{ex post} evaluation, in turn, counts violations according to the evaluator's default bounds, $0.95\leq V\leq1.05$~pu. Consequently, violations in the subsequent reports do not indicate an inconsistency with MISOCP feasibility; they result from a stricter electrical check applied after topology selection. Since current capacity data are not uniformly available for all networks, overload metrics are not used in the main comparison.

\subsection{Test instances}
\label{subsec:instancias_experimentais}

A total of 12 networks were evaluated, ranging from 17 to 10\,561 nodes and from 16 to 10\,728 physical branches. Table~\ref{tab:instancias_v08} summarizes the characteristics provided for each instance. The ``Open sw.'' column corresponds to the number of initially open switches reported in the results database.

\begin{table}[htbp]
\centering
\caption{Characteristics of the instances used in the experiments.}
\label{tab:instancias_v08}
\begin{tabular}{lrrrr}
\hline
Network & Nodes & Branches & Substations & Open sw. \\
\hline
\texttt{bus\_13\_3} & 17 & 16 & 3 & 3 \\
\texttt{bus\_32\_1} & 33 & 37 & 5 & 1 \\
\texttt{bus\_83\_11} & 95 & 96 & 13 & 11 \\
\texttt{bus\_135\_8} & 144 & 156 & 21 & 8 \\
\texttt{bus\_201\_3} & 205 & 216 & 15 & 3 \\
\texttt{bus\_873\_7} & 881 & 900 & 27 & 7 \\
\texttt{bus\_1036\_7} & 1044 & 1178 & 142 & 7 \\
\texttt{bus\_1579\_13} & 1593 & 1798 & 219 & 13 \\
\texttt{bus\_2011\_15} & 2027 & 2289 & 278 & 15 \\
\texttt{bus\_2572\_22} & 2595 & 2931 & 359 & 22 \\
\texttt{bus\_3115\_24} & 3140 & 3546 & 431 & 24 \\
\texttt{bus\_10476\_84} & 10561 & 10728 & 252 & 84 \\
\hline
\end{tabular}
\end{table}

The set allows the method's behavior to be examined from small networks, for which the MISOCP is solved almost instantaneously, to networks with thousands of buses and hundreds of effective topology decisions.

\subsection{Demand scenarios}
\label{subsec:cenarios_experimentais}

Demand uncertainty was represented by three scenarios common to all instances:
\begin{equation}
\begin{array}{c|c|c}
\text{scenario} & \pi_s & \alpha_s \\
\hline
\text{low\_load} & 0.20 & 0.70 \\
\text{nominal\_load} & 0.50 & 1.00 \\
\text{high\_load} & 0.30 & 1.30
\end{array}
\label{eq:exp_scenarios}
\end{equation}
where $\alpha_s$ simultaneously multiplies the active and reactive power demands at all buses:
\begin{equation}
p_i^s=\alpha_s p_i^0,
\qquad
q_i^s=\alpha_s q_i^0.
\label{eq:global_load_scaling}
\end{equation}

The mean multiplier is
\begin{equation}
\bar\alpha
=\sum_{s\in\Omega}\pi_s\alpha_s
=0.2(0.7)+0.5(1.0)+0.3(1.3)
=1.03.
\label{eq:mean_multiplier}
\end{equation}
Thus, the expected-value problem used to calculate EEV is solved with mean loads equivalent to $103\%$ of nominal load. This problem should not be confused with deterministic optimization in the nominal scenario, whose load is exactly $100\%$ of the reference value.

The scenarios constitute a controlled experiment involving global variation in the loading level. They preserve the relative spatial distribution of loads and the power factor at each bus. This feature will be important for interpreting the VSS values and the similarity between deterministic and stochastic topologies.

\subsection{Experimental protocol and metrics}
\label{subsec:protocolo_experimental}

For each network, the initial topology was first evaluated in the three scenarios. The stochastic MISOCP was then solved with a single first-stage topology and scenario-specific operational variables. The resulting topology was fixed and reevaluated under all scenarios. The reports record, among other quantities, expected losses, maximum loss across scenarios, minimum voltage, number of voltage violations, number of switching operations, solver status, \textit{MIP gap}, and solution time.

The expected losses of a topology $y$ are calculated as
\begin{equation}
\mathbb{E}[L(y)]
=\sum_{s\in\Omega}\pi_s L_s(y),
\label{eq:expected_loss_exp}
\end{equation}
and the percentage reduction relative to the initial topology is
\begin{equation}
\Delta L(\%)
=100\frac{\mathbb{E}[L(y^0)]-\mathbb{E}[L(y)]}
{\mathbb{E}[L(y^0)]}.
\label{eq:expected_loss_reduction_exp}
\end{equation}
The worst-case scenario loss is also used:
\begin{equation}
L^{\max}(y)=\max_{s\in\Omega}L_s(y).
\label{eq:worst_loss_exp}
\end{equation}

To quantify the value of the stochastic representation, the canonical definition is used:
\begin{equation}
VSS=EEV-RP,
\label{eq:vss_exp}
\end{equation}
where $RP$ is the expected loss of the topology produced by the stochastic problem, and $EEV$ is the expected loss obtained by fixing, in the original scenarios, the topology selected by the deterministic expected-value problem. In this study, the EV problem uses the probability-weighted mean loads at each bus, corresponding to $\bar\alpha=1.03$.

In addition, the topology comparison files evaluate, under the same scenario distribution, the topologies obtained by deterministic optimization at nominal and peak load. These comparisons are treated separately from VSS because the nominal deterministic topology does not, in general, coincide with the solution of the EV problem.

\subsection{Overall loss results}
\label{subsec:resultados_globais}

Table~\ref{tab:det_stoch_losses_v08} presents the deterministic results previously obtained at the nominal operating point and the stochastic results from the final set of experiments. Deterministic losses are measured in the nominal scenario, whereas stochastic losses are expected values; therefore, the absolute values in the two parts of the table should not be compared directly. The relevant comparison is the percentage reduction achieved by reconfiguration under each criterion.

\begin{table}[htbp]
\centering
\caption{Deterministic losses at the nominal operating point and expected losses in the stochastic formulation.}
\label{tab:det_stoch_losses_v08}
\scriptsize
\setlength{\tabcolsep}{3.5pt}
\resizebox{0.75\textwidth}{!}{%
\begin{tabular}{lrrrrrr}
\hline
& \multicolumn{3}{c}{Deterministic --- nominal load} & \multicolumn{3}{c}{Stochastic --- expected value} \\
Network & Base (kW) & Opt. (kW) & Red. (\%) & Base (kW) & Opt. (kW) & Red. (\%) \\
\hline
\texttt{bus\_13\_3} & 613.25 & 593.44 & 3.23 & 677.64 & 655.75 & 3.23 \\
\texttt{bus\_32\_1} & 182.19 & 130.36 & 28.45 & 201.32 & 144.05 & 28.45 \\
\texttt{bus\_83\_11} & 488.64 & 441.88 & 9.57 & 539.95 & 488.28 & 9.57 \\
\texttt{bus\_135\_8} & 297.21 & 281.75 & 5.20 & 328.41 & 311.33 & 5.20 \\
\texttt{bus\_201\_3} & 511.15 & 486.77 & 4.77 & 564.82 & 537.88 & 4.77 \\
\texttt{bus\_873\_7} & 1413.59 & 1302.46 & 7.86 & 1562.02 & 1439.22 & 7.86 \\
\texttt{bus\_1036\_7} & 1979.38 & 748.42 & 62.19 & 2187.21 & 827.01 & 62.19 \\
\texttt{bus\_1579\_13} & 2467.04 & 969.04 & 60.72 & 2726.08 & 1071.21 & 60.71 \\
\texttt{bus\_2011\_15} & 3240.21 & 1176.93 & 63.68 & 3580.43 & 1300.46 & 63.68 \\
\texttt{bus\_2572\_22} & 3599.20 & 1431.00 & 60.24 & 3977.12 & 1581.26 & 60.24 \\
\texttt{bus\_3115\_24} & 5567.09 & 1935.60 & 65.23 & 6151.64 & 2138.84 & 65.23 \\
\texttt{bus\_10476\_84} & 16585.23 & 14918.00 & 10.05 & 18326.68 & 16484.55 & 10.05 \\
\hline
\end{tabular}%
}
\end{table}

Reconfiguration reduced expected losses in all 12 instances. The percentage reduction ranged from $3.23\%$ in \texttt{bus\_13\_3} to $65.23\%$ in \texttt{bus\_3115\_24}, with a mean of approximately $31.76\%$ and a median of $19.25\%$. The largest relative reductions occurred in \texttt{bus\_1036\_7}, \texttt{bus\_1579\_13}, \texttt{bus\_2011\_15}, \texttt{bus\_2572\_22}, and \texttt{bus\_3115\_24}, all close to or above $60\%$.

A striking result is that the deterministic and stochastic percentage reductions coincide, to the reported precision, in almost all networks; the only visible difference occurs in \texttt{bus\_1579\_13}, whose stochastic solution reached the time limit. This behavior is directly related to the construction of scenarios through global load scaling and is analyzed below.

\subsection{Electrical and operational performance}
\label{subsec:desempenho_eletrico}

Table~\ref{tab:electrical_solver_v08} complements the analysis with the maximum loss across scenarios, the overall minimum voltage, the total number of voltage violations accumulated over the three scenarios, the number of switching operations, and solver indicators.

\begin{table}[htbp]
\centering
\caption{Electrical and computational performance of the stochastic solution. Arrows indicate values before and after reconfiguration.}
\label{tab:electrical_solver_v08}
\scriptsize
\setlength{\tabcolsep}{3pt}
\resizebox{0.8\textwidth}{!}{%
\begin{tabular}{lrrrrrrr}
\hline
Network & $L^{\max}$ (kW) & $V_{\min}$ (pu) & Viol. & Sw. ops. & Status & Gap & Time (s) \\
\hline
\texttt{bus\_13\_3} & 1036.39$\to$1002.92 & 0.939$\to$0.941 & 3$\to$2 & 2 & OPTIMAL & 0.0000 & 0.09 \\
\texttt{bus\_32\_1} & 307.90$\to$220.31 & 0.877$\to$0.915 & 53$\to$18 & 6 & OPTIMAL & 0.0000 & 0.12 \\
\texttt{bus\_83\_11} & 825.81$\to$746.77 & 0.910$\to$0.930 & 28$\to$24 & 14 & OPTIMAL & 0.0000 & 0.03 \\
\texttt{bus\_135\_8} & 502.28$\to$476.15 & 0.914$\to$0.925 & 30$\to$21 & 16 & OPTIMAL & 0.0000 & 0.15 \\
\texttt{bus\_201\_3} & 863.84$\to$822.64 & 0.946$\to$0.948 & 23$\to$6 & 22 & OPTIMAL & 0.0000 & 0.13 \\
\texttt{bus\_873\_7} & 2388.97$\to$2201.16 & 0.944$\to$0.946 & 122$\to$98 & 36 & OPTIMAL & 0.0001 & 1.45 \\
\texttt{bus\_1036\_7} & 3345.15$\to$1264.85 & 0.947$\to$0.979 & 62$\to$0 & 164 & OPTIMAL & 0.0000 & 5.38 \\
\texttt{bus\_1579\_13} & 4169.31$\to$1638.32 & 0.944$\to$0.974 & 122$\to$0 & 228 & TIME\_LIMIT & 0.0243 & 600.01 \\
\texttt{bus\_2011\_15} & 5475.95$\to$1988.94 & 0.947$\to$0.989 & 62$\to$0 & 300 & OPTIMAL & 0.0001 & 317.81 \\
\texttt{bus\_2572\_22} & 6082.65$\to$2418.40 & 0.942$\to$0.984 & 135$\to$0 & 382 & OPTIMAL & 0.0001 & 143.21 \\
\texttt{bus\_3115\_24} & 9408.39$\to$3271.16 & 0.944$\to$0.959 & 345$\to$0 & 472 & OPTIMAL & 0.0001 & 98.93 \\
\texttt{bus\_10476\_84} & 28029.04$\to$25211.67 & 0.942$\to$0.943 & 604$\to$545 & 332 & OPTIMAL & 0.0001 & 33.61 \\
\hline
\end{tabular}%
}
\end{table}

The maximum loss was reduced in every instance. The same holds for the number of voltage violations. Across the 12 networks and three scenarios, the total number of violations fell from 1589 to 714, a reduction of approximately $55.1\%$. Five networks --- \texttt{bus\_1036\_7}, \texttt{bus\_1579\_13}, \texttt{bus\_2011\_15}, \texttt{bus\_2572\_22}, and \texttt{bus\_3115\_24} --- achieved zero violations in the \textit{ex post} evaluation.

Loss reduction was not accompanied by explicit minimization of the number of switching operations because $\gamma=0$. Thus, the ``Sw. ops.'' column should be interpreted as a characteristic of the selected topology. The number of operations increases substantially in some larger networks, reaching 472 in \texttt{bus\_3115\_24}. This result reinforces the usefulness of the optional switching penalty term for future experiments, without changing the analysis of the current experiments.

\subsection{Value of the stochastic solution}
\label{subsec:vss_resultados}

Table~\ref{tab:vss_v08} presents $RP$, $EEV$, and $VSS$ calculated using the expected-value procedure in the main report. The ``Same top.'' column indicates whether the topology of the stochastic problem exactly matches that produced by the EV problem.

\begin{table}[htbp]
\centering
\caption{Value of the stochastic solution. $EEV$ is obtained by solving the deterministic problem at the probability-weighted mean loads and then evaluating the solution in the original scenarios.}
\label{tab:vss_v08}
\scriptsize
\setlength{\tabcolsep}{3pt}
\resizebox{0.75\textwidth}{!}{%
\begin{tabular}{lrrrrlll}
\hline
Network & RP (kW) & EEV (kW) & VSS (kW) & VSS (\%) & Same top. & RP status & EV status \\
\hline
\texttt{bus\_13\_3} & 655.753 & 655.753 & 0.0000 & 0.0000 & yes & OPTIMAL & OPTIMAL \\
\texttt{bus\_32\_1} & 144.047 & 144.047 & 0.0000 & 0.0000 & yes & OPTIMAL & OPTIMAL \\
\texttt{bus\_83\_11} & 488.275 & 488.275 & 0.0000 & 0.0000 & yes & OPTIMAL & OPTIMAL \\
\texttt{bus\_135\_8} & 311.332 & 311.332 & 0.0000 & 0.0000 & yes & OPTIMAL & OPTIMAL \\
\texttt{bus\_201\_3} & 537.881 & 537.881 & 0.0000 & 0.0000 & yes & OPTIMAL & OPTIMAL \\
\texttt{bus\_873\_7} & 1439.220 & 1439.220 & 0.0000 & 0.0000 & yes & OPTIMAL & OPTIMAL \\
\texttt{bus\_1036\_7} & 827.015 & 827.001 & -0.0138 & -0.0017 & no & OPTIMAL & OPTIMAL \\
\texttt{bus\_1579\_13} & 1071.206 & 1071.585 & 0.3786 & 0.0353 & no & TIME\_LIMIT & TIME\_LIMIT \\
\texttt{bus\_2011\_15} & 1300.458 & 1300.642 & 0.1848 & 0.0142 & no & OPTIMAL & TIME\_LIMIT \\
\texttt{bus\_2572\_22} & 1581.261 & 1581.261 & 0.0000 & 0.0000 & yes & OPTIMAL & OPTIMAL \\
\texttt{bus\_3115\_24} & 2138.837 & 2138.855 & 0.0182 & 0.0008 & no & OPTIMAL & OPTIMAL \\
\texttt{bus\_10476\_84} & 16484.553 & 16484.456 & -0.0962 & -0.0006 & no & OPTIMAL & OPTIMAL \\
\hline
\end{tabular}%
}
\end{table}

In seven instances, the RP and EV topologies are identical and $VSS=0$. In the other five, the topologies differ, but the performance difference remains extremely small. Among the ten instances in which both RP and EV reached \texttt{OPTIMAL} status, the largest absolute relative VSS did not exceed $0.0017\%$. The two exceptions regarding optimality certification are \texttt{bus\_1579\_13}, in which both problems reached the time limit, and \texttt{bus\_2011\_15}, in which the EV problem reached the time limit.

The small negative values observed in \texttt{bus\_1036\_7} and \texttt{bus\_10476\_84} should not be interpreted as a negative economic value of stochastic information. The theoretical VSS is nonnegative when RP and EV are solved exactly and evaluated using the same objective function. In this experiment, however, the reported value is calculated using losses from the \textit{ex post} evaluation, whereas the MISOCP uses the approximate conic representation and is solved subject to numerical tolerances. Differences on the order of hundredths of a kW in networks with losses of hundreds or tens of thousands of kW are therefore treated as numerical equivalence.

\subsection{Comparison of deterministic and stochastic topologies}
\label{subsec:comparacao_topologias}

The comparison reports also solve two deterministic benchmarks: one topology optimized for nominal load and another for peak load. The four topologies --- initial, nominal deterministic, peak deterministic, and stochastic --- are then evaluated under the same three scenarios.

In the six smallest instances, from \texttt{bus\_13\_3} to \texttt{bus\_873\_7}, the nominal deterministic, peak deterministic, and stochastic topologies coincided exactly. The same occurred in \texttt{bus\_2572\_22}. In the remaining five networks, different topologies were obtained, but the differences in expected loss remained small. The largest gain from the stochastic topology relative to the nominal deterministic topology occurred in \texttt{bus\_3115\_24}: expected loss decreased from 2143.146 to 2138.837~kW, an additional reduction of approximately $0.201\%$. In \texttt{bus\_2011\_15}, the gain was approximately $0.040\%$. In the other instances with different topologies, the variations were below $0.04\%$ in magnitude or occurred in runs terminated by the time limit.

This comparison confirms that, for the adopted scenario set, the main contribution of optimization lies in reconfiguring the network relative to its initial topology; the additional benefit of explicitly considering the three scenarios, compared with a well-chosen deterministic topology, is small.

\subsection{Scenario-specific behavior}
\label{subsec:comportamento_cenarios}

All 36 power flow calculations corresponding to the 12 stochastic topologies and three scenarios converged, and all evaluated configurations were classified as radial. In every network, \texttt{high\_load} was both the scenario with the highest loss and the scenario with the lowest voltage. This result is consistent with the fact that the three scenarios differ only by a global load multiplier.

Voltage violations are strongly concentrated in the highest-load scenario. Before reconfiguration, the total number of violations accumulated across the 12 networks was 13 in \texttt{low\_load}, 40 in \texttt{nominal\_load}, and 1536 in \texttt{high\_load}. After reconfiguration, these values became 0, 19, and 695, respectively. At low load, only one network initially had violations, and none retained violations after reconfiguration. In the peak-load scenario, all 12 networks initially had at least one violation, while seven still had some violations after optimization.

The equality of percentage loss reductions across the three scenarios can be explained by the structure of the experiment. For a fixed topology, the adopted approximation yields flows proportional to the multiplier $\alpha_s$ and losses that are quadratic in the flows. Thus,
\begin{equation}
L_s(y)\approx \alpha_s^2 L_{\mathrm{nom}}(y),
\label{eq:loss_scaling_exp}
\end{equation}
and therefore,
\begin{equation}
\mathbb{E}[L(y)]
\approx
\left(\sum_{s\in\Omega}\pi_s\alpha_s^2\right)
L_{\mathrm{nom}}(y)
=1.105\,L_{\mathrm{nom}}(y).
\label{eq:expected_loss_scaling_exp}
\end{equation}
The factor $1.105$ is independent of the topology. When voltage constraints do not change the set of competitive topologies, minimizing nominal loss or expected loss therefore yields the same ordering of configurations. This property explains the observed coincidence of percentage reductions and the nearly zero VSS.

\subsection{Computational performance and scalability}
\label{subsec:desempenho_computacional}

Eleven of the 12 instances terminated with \texttt{OPTIMAL} status and a \textit{MIP gap} consistent with the $10^{-4}$ tolerance. The only exception was \texttt{bus\_1579\_13}, which terminated at the 600~s limit with a \textit{gap} of approximately $2.43\%$. Among the 11 instances with an optimality certificate, the median time was approximately 1.45~s. Seven were solved in less than 10~s and nine in less than 100~s.

Solution time does not increase monotonically with the number of buses. Network \texttt{bus\_2011\_15}, with 2027 nodes, required approximately 318~s, whereas \texttt{bus\_10476\_84}, with 10\,561 nodes, was solved in about 34~s. This behavior shows that computational effort depends not only on network size but also on the structure of the candidate branches, relaxation quality, the location of alternative topology choices, and the interaction among the \textit{warm start}, cycle cuts, and integer search.

Instance \texttt{bus\_1579\_13} illustrates the main limitation observed in this set of experiments. Although a high-quality solution was found --- with an expected loss reduction of approximately $60.71\%$ and elimination of voltage violations in the subsequent evaluation --- global optimality was not certified within the prescribed limit. This case motivates future work on stronger relaxations and decomposition techniques, while still allowing the current results to be used to assess the performance of an incumbent solution with a known \textit{gap}.

\subsection{Discussion of results}
\label{subsec:discussao_resultados}

The experiments allow two effects to be distinguished. The first is the value of reconfiguration itself: in every network, changing the topology reduced expected losses and the maximum observed loss, while improving or preserving the minimum voltage. This effect is substantial in several instances, with loss reductions exceeding $60\%$.

The second effect is the incremental value of stochastic modeling. In these experiments, this effect was small. The scenarios retain the same spatial load distribution and vary only the overall loading level. Since the loss function in the adopted approximation essentially preserves a common quadratic scaling structure across scenarios, topologies that minimize nominal loss also tend to minimize expected loss. The VSS analysis confirms this behavior: seven networks have identical RP--EV topologies, and performance differences in the remaining networks are numerically small.

This result does not constitute evidence that stochastic programming is unhelpful; it characterizes a property of the scenario set used as a controlled \textit{baseline}. For stochastic information to have greater potential to change the first-stage decision, it is necessary to introduce scenarios that alter the spatial distribution of demand, correlations among network regions, distinct active and reactive power variations, or other sources of uncertainty. The current experiments therefore serve two purposes: they computationally validate the multiscenario formulation on networks of varying sizes and quantitatively identify the limitation of scenarios obtained solely through global scaling.

In summary, the approach demonstrated the ability to produce topologies with substantially lower losses and improved voltage behavior, while providing either an exact solution or an explicitly known \textit{gap}. At the same time, VSS values close to zero indicate that the next methodological step should focus on constructing spatially heterogeneous scenarios and systematically analyzing the trade-off among losses, risk, and the number of switching operations.

\section{Concluding Remarks}
\label{sec:consideracoes_finais}

The results discussed in the previous section bring us back to the central objective of this study: to evaluate stochastic distribution network reconfiguration under demand uncertainty in controlled conditions, separating the effect of a topology change from the additional effect of explicitly considering scenarios. To this end, a two-stage stochastic model was formulated in which topology decisions are made in the first stage and remain common to all scenarios, whereas flows, voltages, and auxiliary variables are determined in the second stage. The deterministic equivalent was represented as a multiscenario MISOCP with explicit radiality, a LinDistFlow approximation, a convex quadratic representation of losses, and optional switching penalties. Computationally, the model was solved monolithically with Gurobi, using a \textit{warm start} and dynamic cycle separation through \textit{lazy constraints}.

Experiments on 12 networks, ranging from 17 to 10\,561 nodes, showed that reconfiguration reduced expected losses in every instance. Reductions ranged from $3.23\%$ to $65.23\%$, averaging approximately $31.76\%$. The maximum loss observed across scenarios was also reduced in every network, and the cumulative number of voltage violations in the \textit{ex post} evaluation decreased from 1589 to 714, a reduction of approximately $55.1\%$. Eleven of the 12 instances were solved with \texttt{OPTIMAL} status; only \texttt{bus\_1579\_13} reached the 600~s limit, with a \textit{gap} of approximately $2.43\%$. These results indicate that the adopted formulation can produce valid radial topologies with consistent improvements in electrical performance across networks of different sizes, while providing either an optimality certificate or explicit information about the attained \textit{gap}.

The analysis of the value of the stochastic solution yields an important additional conclusion. In seven networks, the RP and EV problems produced exactly the same topology and $VSS=0$; in the remaining networks, performance differences were very small. This behavior stems from the structure of the scenarios used, which change the global loading level by factors of $0.70$, $1.00$, and $1.30$, preserving the relative spatial distribution of active and reactive power demands. Under the adopted loss approximation, this scaling tends to preserve the ordering of candidate topologies. Thus, the results show that the main gain in these experiments is associated with reconfiguring the network relative to its initial topology, whereas the incremental value of stochastic information is small when uncertainty is represented solely through global, homogeneous load scaling. This finding is a methodological result of the study because it delineates the conditions under which introducing scenarios may or may not change the first-stage decision.

This perspective complements the studies reviewed in the literature. Two-stage robust formulations, such as that of Lee et al.~\cite{lee2015}, and data-driven stochastic--robust approaches, such as that of Akrami et al.~\cite{akrami2020}, focus on protection against uncertainty sets and adverse realizations. Mahdavi, Schmitt, and Jurado~\cite{mahdavi2023} present a monolithic MISOCP formulation for simultaneous uncertainty in load and distributed generation, whereas Esmaeilnezhad et al.~\cite{esmaeilnezhad2025} and Yahia and Gheith~\cite{yahia2026} extend the problem to multiple sources of uncertainty, dynamic operation, or microgrids. The present study does not seek to replace these approaches or simultaneously increase the number of technologies modeled. Its main scientific contribution is to establish a two-stage stochastic formulation and a controlled experimental protocol that allow the benefit of reconfiguration and the value of stochastic information to be identified separately, using metrics such as RP, EEV, and VSS, alongside electrical and computational indicators evaluated under the same conditions.

From an academic perspective, this structure provides a reference framework for subsequent extensions and comparisons among different representations of uncertainty, preserving traceability between the model, implementation, and results. From a practical perspective, the experiments show that appropriate topology selection can reduce losses and improve the voltage profile even when demand varies across scenarios. The formulation also retains the option to penalize the number or cost of switching operations, although this term was disabled in the reported results. In real applications, this feature allows operational restrictions associated with crew deployment, equipment wear, or operating policies to be incorporated. In terms of social impact, reducing technical losses and improving the voltage profile may contribute to more efficient use of electrical infrastructure and better supply quality, although economic and social quantification of these effects is outside the scope of this study.

The conclusions should be interpreted in light of several limitations. Uncertainty was restricted to demand and represented by only three global-scaling scenarios, without spatial heterogeneity, temporal dependence, or calibration from observed probabilistic data. No out-of-sample evaluation was performed. The electrical model and the \textit{ex post} evaluation use approximations based on DistFlow/LinDistFlow and therefore do not replace higher-fidelity nonlinear AC validation. Moreover, the voltage bounds used in optimization are more permissive than the range used in the subsequent evaluation, so the absence of violations in the $0.95$--$1.05$~pu range is not directly enforced by the MISOCP. Branch current capacities are also not uniformly available across the instances, which is why overloads were not used as a primary metric. Finally, switching penalties were retained in the general model, but $\gamma=0$ was adopted in the current experiments.

As a first extension, future work will construct spatially heterogeneous scenarios with correlations across buses, preferably calibrated from historical data, and evaluate solutions through out-of-sample procedures. Sample average approximation, scenario reduction, and risk measures such as CVaR may extend the analysis of uncertainty effects. Studying different values of $\gamma$ and switching costs is also relevant for characterizing the trade-off between expected losses and operational effort. Other extensions include distributed generation, storage, capacitor banks, and multiperiod models; electrical validation using AC power flow or higher-fidelity radial methods; systematic incorporation of thermal limits; and comparisons of stochastic programming, robust optimization, and distributionally robust optimization using the same set of instances. For instances in which the monolithic solution approach becomes a limitation, decomposition techniques, stronger relaxations, and valid inequalities are natural alternatives.

In summary, this study achieved its objective of establishing and evaluating a controlled stochastic framework for distribution network reconfiguration under demand uncertainty. The results confirm the benefit of reconfiguration in reducing losses and improving voltage behavior, while also showing that homogeneous loading scenarios can yield a nearly zero incremental stochastic value. This distinction between the value of reconfiguration and the value of stochastic information is the main conclusion of the study and provides an objective reference for developing future models with more representative scenarios, greater electrical fidelity, and additional operational criteria.

\section*{Acknowledgments}
This work was supported by the Brazilian National Council for Scientific and Technological Development (CNPq) (grant numbers 351735/2025-9 and 311907/2023-7).

\printbibliography[title={References}] 

@techreport{antmann2009,
  author      = {Antmann, Pedro},
  title       = {Reducing Technical and Non-Technical Losses in the Power Sector},
  institution = {World Bank Group},
  year        = {2009},
  month       = {July},
  type        = {Background Paper for the World Bank Group Energy Sector Strategy},
  address     = {Washington, DC},
  url         = {http://worldbank.org},
  note        = {Public Disclosure Authorized}
}

@techreport{aneel2025,
  author      = {{Agência Nacional de Energia Elétrica}},
  title       = {Perdas de Energia Elétrica na Distribuição: 2025/2024},
  institution = {ANEEL, Superintendência de Gestão Tarifária e Regulação Econômica (STR)},
  year        = {2025},
  month       = {maio},
  type        = {Relatório Técnico},
  address     = {Brasília, DF},
  url         = {https://www.gov.br},
  note        = {Elaborado pela Coordenação de Monitoramento Tarifário e Avaliação Regulatória e Coordenação de Regulação Tarifária}
}

@online{epe-2026,
  author       = {{Empresa de Pesquisa Energ\'etica (EPE)}},
  title        = {EPE atualiza painel interativo sobre MMGD com dados de 2025},
  year         = {2026},
  url          = {https://www.epe.gov.br/pt/imprensa/noticias/epe-atualiza-painel-interativo-sobre-mmgd-com-dados-de-2025},
  urldate      = {2026-09-03},
  langid       = {portuguese},
}

@article{civanlar1988,
  author       = {Civanlar, S. and Grainger, J. J. and Yin, H. and Lee, S. S. H.},
  title        = {Distribution feeder reconfiguration for loss reduction},
  journal      = {IEEE Transactions on Power Delivery},
  volume       = {3},
  number       = {3},
  pages        = {1217--1223},
  year         = {1988},
  doi          = {10.1109/61.193882},
}

@article{baran1989,
  author       = {Baran, M. E. and Wu, F. F.},
  title        = {Network reconfiguration in distribution systems for loss reduction and load balancing},
  journal      = {IEEE Transactions on Power Delivery},
  volume       = {4},
  number       = {2},
  pages        = {1401--1407},
  year         = {1989},
  doi          = {10.1109/61.25627},
}

@article{baran1989b,
  author    = {Baran, Mesut E. and Wu, Felix F.},
  journal   = {IEEE Transactions on Power Delivery},
  title     = {Optimal capacitor placement on radial distribution systems},
  year      = {1989},
  volume    = {4},
  number    = {1},
  pages     = {725--734},
  doi       = {10.1109/61.19265},
  publisher = {IEEE}
}

@article{shirmohammadi1989,
  author    = {Shirmohammadi, Dariush and Hong, H. Wayne},
  title     = {Reconfiguration of electric distribution networks for resistive line losses reduction},
  journal   = {IEEE Transactions on Power Delivery},
  volume    = {4},
  number    = {2},
  pages     = {1492--1498},
  year      = {1989},
  month     = {April},
  doi       = {10.1109/61.25627},
  publisher = {IEEE}
}

@article{taylor1990,
  author    = {Taylor, Tim and Lubkeman, David},
  journal   = {IEEE Transactions on Power Delivery},
  title     = {Implementation of heuristic search strategies for distribution feeder reconfiguration},
  year      = {1990},
  volume    = {5},
  number    = {1},
  pages     = {239--246},
  doi       = {10.1109/61.107277},
  publisher = {IEEE}
}

@article{wagner1991,
  author    = {Wagner, T. P. and Chikhani, A. Y. and Hackam, R.},
  title     = {Feeder reconfiguration for loss reduction: An application of distribution automation},
  journal   = {IEEE Transactions on Power Delivery},
  volume    = {6},
  number    = {4},
  pages     = {1922--1933},
  year      = {1991},
  month     = {October},
  doi       = {10.1109/61.97707},
  publisher = {IEEE}
}

@article{goswami1992,
  author    = {Goswami, S.-K. and Basu, S. K.},
  title     = {A new algorithm for the reconfiguration of distribution feeders for loss minimization},
  journal   = {IEEE Transactions on Power Delivery},
  volume    = {7},
  number    = {3},
  pages     = {1484--1491},
  year      = {1992},
  month     = {July},
  doi       = {10.1109/61.141868},
  publisher = {IEEE}
}

@article{chen1993,
  author    = {Chen, C. S. and Cho, M. Y.},
  title     = {Energy loss reduction by critical switches},
  journal   = {IEEE Transactions on Power Delivery},
  volume    = {8},
  number    = {3},
  pages     = {1246--1253},
  year      = {1993},
  month     = {July},
  doi       = {10.1109/61.252656},
  publisher = {IEEE}
}

@CONFERENCE {Merlin1975,
    author    = "Merlin, A. and Back, H",
    title     = "Search for a Minimal-Loss Operating Spanning Tree Configuration in Urban Power Distribution Systems",
    booktitle = "5th Power Systems Computation Conference",
    year      = "1975",
    pages     = "1-5",
    address   = "Cambridge, U.K"
}

@article{cavellucci1997,
  title={Minimization of energy losses in electric power distribution systems by intelligent search strategies},
  author={Cavellucci, Celso and Lyra, Christiano},
  journal={International Transactions in Operational Research},
  volume={4},
  number={1},
  pages={23--33},
  year={1997},
  publisher={Elsevier}
}

@INPROCEEDINGS{Jeon2000,
  author={Young-Jae Jeon and Jae-Chul Kim},
  booktitle={2000 IEEE Power Engineering Society Winter Meeting. Conference Proceedings (Cat. No.00CH37077)}, 
  title={Network reconfiguration in radial distribution system using simulated annealing and Tabu search}, 
  year={2000},
  volume={4},
  number={},
  pages={2329-2333 vol.4},
  doi={10.1109/PESW.2000.847169}}

@inproceedings{lyra2003,
    author = {Lyra, Christiano and Cristiane P. Fernandes and Celso Cavellucci},
    title ={Informed Searches Uncover Paths to Enhance Energy Output in Power Distribution Networks} ,
    booktitle = {Proceedings of the "12th Intelligent  Systems Application to Power Systems Conference},
    year = {2003}
}

@inproceedings{bueno2004,
  author    = {Bueno, Edilson A. and Lyra Filho, Christiano and Cavellucci, Celso},
  title     = {Redu{\c{c}}{\~a}o de perdas na distribu{\c{c}}{\~a}o de energia el{\'e}trica pelo m{\'e}todo {GRASP}},
  booktitle = {Anais do XV Congresso Brasileiro de Autom{\'a}tica (CBA 2004)},
  address   = {Campinas, SP, Brasil},
  pages     = {1253--1258},
  year      = {2004},
  publisher = {SBA}
}

@ARTICLE{Enacheanu2008,
  author={Enacheanu, Bogdan and Raison, Bertrand and Caire, Raphael and Devaux, Olivier and Bienia, Wojciech and HadjSaid, Nouredine},
  journal={IEEE Transactions on Power Systems}, 
  title={Radial Network Reconfiguration Using Genetic Algorithm Based on the Matroid Theory}, 
  year={2008},
  volume={23},
  number={1},
  pages={186-195},
  doi={10.1109/TPWRS.2007.913303}}

@article{queiroz2009,
  title={Adaptive hybrid genetic algorithm for technical loss reduction in distribution networks under variable demands},
  author={Queiroz, Leonardo MO and Lyra, Christiano},
  journal={IEEE Transactions on Power Systems},
  volume={24},
  number={1},
  pages={445--453},
  year={2009},
  publisher={IEEE}
}

@inproceedings{torres2010,
  title={Reconfiguration of power distribution systems using genetic algorithms and spanning trees},
  author={Torres-Jimenez, J and Guardado, JL and Rivas, F and Maximov, S and Melgoza, E},
  booktitle={2010 IEEE Electronics, Robotics and Automotive Mechanics Conference},
  pages={779--784},
  year={2010},
  organization={IEEE}
}

@Article{Jakus2020,
AUTHOR = {Jakus, Damir and Čađenović, Rade and Vasilj, Josip and Sarajčev, Petar},
TITLE = {Optimal Reconfiguration of Distribution Networks Using Hybrid Heuristic-Genetic Algorithm},
JOURNAL = {Energies},
VOLUME = {13},
YEAR = {2020},
NUMBER = {7},
ARTICLE-NUMBER = {1544},
URL = {https://www.mdpi.com/1996-1073/13/7/1544},
ISSN = {1996-1073},
DOI = {10.3390/en13071544}
}

@article{jabr2012,
  author    = {Jabr, Rabih A. and Singh, Ravindra and Pal, Bikash C.},
  title     = {Minimum Loss Network Reconfiguration Using Mixed-Integer Convex Programming},
  journal   = {IEEE Transactions on Power Systems},
  volume    = {27},
  number    = {2},
  pages     = {1106--1115},
  year      = {2012},
  publisher = {IEEE},
  doi       = {10.1109/TPWRS.2011.2180406}
}

@article{taylor2012,
  author    = {Taylor, Joshua A. and Hover, Franz S.},
  title     = {Convex Models of Distribution System Reconfiguration},
  journal   = {IEEE Transactions on Power Systems},
  volume    = {27},
  number    = {3},
  pages     = {1407--1413},
  year      = {2012},
  publisher = {IEEE},
}

@inproceedings{mahdavi2019,
  author    = {Mahdavi, Meisam and Macedo, Leonardo H. and Romero, Rub{\'e}n},
  title     = {A Mathematical Formulation for Distribution Network Reconfiguration},
  booktitle = {Proceedings of the 7th Regional Conference on Electricity Distribution (CRTS)},
  address   = {Tehran, Iran},
  pages     = {1--5},
  year      = {2019},
  organization = {S{\~a}o Paulo Research Foundation (FAPESP)}
}

@inproceedings{mahdavi2021,
  author    = {Mahdavi, Meisam and Javadi, Mohammad S. and Wang, Fei and Catal{\~a}o, Jo{\~a}o P. S.},
  title     = {Optimal Modeling of Load Variations in Distribution System Reconfiguration},
  booktitle = {Proceedings of the 2021 IEEE International Conference on Environment and Electrical Engineering and 2021 IEEE Industrial and Commercial Power Systems Europe (EEEIC / I\&CPS Europe)},
  address   = {Bari, Italy},
  pages     = {1--6},
  year      = {2021},
  publisher = {IEEE},
}

@article{belotti2025,
  author    = {Belotti, J{\^o}natas Trabuco and Cavellucci, Celso and Siqueira, Hugo Valadares and Usberti, F{\'a}bio Luiz},
  title     = {Evaluation of radiality constraints for power distribution networks},
  journal   = {RAIRO - Operations Research},
  volume    = {59},
  pages     = {595--608},
  year      = {2025},
  publisher = {EDP Sciences},
  doi       = {10.1051/ro/2025003}
}

@article{ayazi2026,
  author    = {Ayazi, Amir Mohammad and Shakarami, Mahmood Reza and Doostizadeh, Meysam and Namdari, Farhad and Nikzad, Mohammad Reza},
  title     = {An {MILP} approach for optimal operation of unbalanced distribution networks through coordinated network reconfiguration and {SOP} utilization},
  journal   = {Scientific Reports},
  volume    = {16},
  number    = {2069},
  pages     = {1--24},
  year      = {2026},
  publisher = {Nature Publishing Group}
}

@article{otuo_acheampong2023,
  author    = {Otuo-Acheampong, Duku and Rashed, Ghamgeen Izat and Akwasi, Amoh Mensah and Haider, Hussain},
  title     = {Application of Optimal Network Reconfiguration for Loss Minimization and Voltage Profile Enhancement of Distribution System Using Heap-Based Optimizer},
  journal   = {International Transactions on Electrical Energy Systems},
  volume    = {2023},
  number    = {Article ID 9930954},
  pages     = {1--17},
  year      = {2023},
  publisher = {Hindawi / Wiley},
  doi       = {10.1155/2023/9930954}
}

@article{Subburaj2006,
author = {Perumal, Subburaj and Ramar, K. and Ganesan, Lakshmanan and Venkatesh, P.},
year = {2006},
month = {12},
pages = {},
title = {Distribution System Reconfiguration for Loss Reduction using Genetic Algorithm},
volume = {2},
journal = {Journal of Electrical Systems}
}

@article{mahendran2021,
  author    = {Mahendran, G. and Govindaraju, C.},
  title     = {Combined phase balancing and reconfiguration for unbalanced distribution system optimization through hybrid fuzzy-MFOA},
  journal   = {Journal of Ambient Intelligence and Humanized Computing},
  volume    = {12},
  number    = {1},
  pages     = {1--10},
  year      = {2021},
  publisher = {Springer},
  doi       = {10.1007/s12652-021-02966-8}
}

@article{kagan1998,
  author    = {Kagan, Nelson and Oliveira, Carlos C{\'e}sar Barioni de},
  title     = {Reconfigura{\c{c}}{\~a}o de redes de distribui{\c{c}}{\~a}o de energia el{\'e}trica atrav{\'e}s de ferramenta para solu{\c{c}}{\~a}o de problemas de decis{\~a}o com m{\'u}ltiplos objetivos e incertezas},
  journal   = {Controle \& Automa{\c{c}}{\~a}o},
  volume    = {9},
  number    = {1},
  pages     = {18--30},
  year      = {1998},
  publisher = {Sociedade Brasileira de Autom{\'a}tica (SBA)}
}

@ARTICLE{Malekpour2013,
  author={Malekpour, Ahmad Reza and Niknam, Taher and Pahwa, Anil and Kavousi Fard, Abdollah},
  journal={IEEE Transactions on Power Systems}, 
  title={Multi-Objective Stochastic Distribution Feeder Reconfiguration in Systems With Wind Power Generators and Fuel Cells Using the Point Estimate Method}, 
  year={2013},
  volume={28},
  number={2},
  pages={1483-1492},
  doi={10.1109/TPWRS.2012.2218261}
}

@article{lee2015,
  author       = {Lee, C. and Liu, C. and Mehrotra, S. and Bie, Z.},
  title        = {Robust Distribution Network Reconfiguration},
  journal      = {IEEE Transactions on Smart Grid},
  volume       = {6},
  number       = {2},
  pages        = {836--842},
  year         = {2015},
  doi          = {10.1109/TSG.2014.2375160},
}

@article{akrami2020,
  author       = {Akrami, A. and Doostizadeh, M. and Aminifar, F.},
  title        = {Optimal Reconfiguration of Distribution Network Using {$\mu$PMU} Measurements: A Data-Driven Stochastic Robust Optimization},
  journal      = {IEEE Transactions on Smart Grid},
  volume       = {11},
  number       = {1},
  pages        = {420--428},
  year         = {2020},
  doi          = {10.1109/TSG.2019.2925700},
}

@article{mahdavi2023,
  author       = {Mahdavi, M. and Schmitt, K. E. K. and Jurado, F.},
  title        = {Robust Distribution Network Reconfiguration in the Presence of Distributed Generation Under Uncertainty in Demand and Load Variations},
  journal      = {IEEE Transactions on Power Delivery},
  volume       = {38},
  number       = {5},
  pages        = {3480--3495},
  year         = {2023},
  doi          = {10.1109/TPWRD.2023.3276634},
}

@article{esmaeilnezhad2025,
  author       = {Esmaeilnezhad, Behzad and Amini, Hossein and Noroozian, Reza and Jalilzadeh, Saeid},
  title        = {Flexible Reconfiguration for Optimal Operation of Distribution Network Under Renewable Generation and Load Uncertainty},
  journal      = {Energies},
  volume       = {18},
  number       = {2},
  pages        = {266},
  year         = {2025},
  doi          = {10.3390/en18020266},
}

@article{yahia2026,
  author       = {Yahia, Z. and Gheith, M.},
  title        = {A stochastic model for dynamic reconfiguration of multi-microgrid networks under demand and supply uncertainties},
  journal      = {Scientific Reports},
  volume       = {16},
  number       = {1},
  pages        = {15489},
  year         = {2026},
  doi          = {10.1038/s41598-026-52537-0},
}

@article{vielma2017,
  author       = {Vielma, Juan Pablo and Ahmed, Shabbir and Nemhauser, George L.},
  title        = {Mixed-Integer Second-Order Cone Programming: A Survey},
  journal      = {INFORMS Tutorials in Operations Research},
  volume       = {2017},
  pages        = {135--156},
  year         = {2017},
  publisher    = {INFORMS},
  doi          = {10.1287/educ.2017.0167},
}

@book{birge2011,
  title        = {Introduction to Stochastic Programming},
  author       = {Birge, John R. and Louveaux, Fran\c{c}ois},
  year         = {2011},
  edition      = {Second},
  publisher    = {Springer},
  series       = {Springer Series in Operations Research and Financial Engineering},
  address      = {New York},
  isbn         = {978-0-387-98217-5},
}

\end{document}